# Effective *Ex-Situ* Fabrication of F-Doped SmFeAsO Wire for High Transport Critical Current Density

**Masaya Fujioka[1*], Tomohiro Kota[1], Masanori Matoba[1], Toshinori Ozaki[2], Yoshihiko Takano[2,3], Hiroaki Kumakura[2,3], and Yoichi Kamihara[1,3]**

[1]Department of Applied Physics and Physico-Informatics, Faculty of Science and Technology, Keio University, 3-14-1 Hiyoshi, Yokohama 223-8522, Japan

[2]National Institute for Materials Science, 1-2-1 Sengen, Tsukuba 305-0047, Japan

[3]TRIP, Japan Science and Technology Agency, Sanban-cho bldg, 5, Sanban-cho, Chiyoda, Tokyo 102-0075, Japan

**Abstract.**

We demonstrate the fabrication of superconducting $SmFeAsO_{1-x}F_x$ (Sm-1111) wires by using the *ex-situ* powder-in-tube technique. Sm-1111 powder and a binder composed of $SmF_3$, samarium arsenide, and iron arsenide were used to synthesize the superconducting core. Although the F content of Sm-1111 is reduced in the process of *ex-situ* fabrication, the binder compensates by sufficiently supplementing the F content, thereby preventing a decrease in the superconducting transition temperature and a shrinkage of the superconducting volume fraction. Thus, in the superconducting Sm-1111 wire with the binder, the transport critical current density reaches the highest value of ~4 kA/cm² at 4.2 K.

[*] Corresponding E-mail address: fujioka-masaya@a6.keio.jp

Iron-based superconductors[1,2] demonstrate a relatively high superconducting transition temperature $T_c$ and very high upper critical magnetic field $H_{c2}$.[3,4] These attractive characteristics of iron-based superconductors triggered research into new iron-based superconductors as well as several attempts at applications. Iron-based superconductors, such as $SmFeAsO_{1-x}F_x$ (Sm-1111), have critical temperatures above 50 K.[2]

For iron-based superconducting wires, such as $(Ba, K)Fe_2As_2$, the transport critical current density $J_c$ attains a value of $\sim 1.0 \times 10^4$ A/cm².[5] These iron-based superconducting wires were fabricated by the powder-in-tube (PIT) technique[6] in which the powders, which are synthesized for the superconducting core, are packed into a metal tube and the tubes are annealed after wire drawing. The PIT technique is classified into two different processes: *in-situ* and *ex-situ*. The former employs powder starting materials, whereas the latter employs a powder of synthesized superconducting material.

Wang *et al.* succeeded in fabricating a superconducting Sm-1111 wire, using silver as a sheath material, and obtained a transport $J_c$ of 1.3 kA/cm² at 4.2 K.[7] Moreover, Ma *et al.* reported that a tape made of superconducting Sm-1111 with a silver sheath demonstrated a transport $J_c$ of as high as 2.7 kA/cm².[8] These Sm-1111 wires were fabricated by the *in-situ* PIT technique;[7-10] however, to date, no report exists detailing the use of the *ex-situ* PIT technique to fabricate Sm-1111 wires.

Thus, in this study, we demonstrate, for the first time, an *ex-situ* PIT technique that uses an Ag sheath and gives improved superconducting properties for Sm-1111 superconducting wires by using a binder that is composed of $SmF_3$, Sm arsenide, and Fe arsenide.

Several polycrystalline F-doped SmFeAsO bulk samples were prepared by a solid-state reaction. Figure 1 shows a flowchart that describes our method of fabricating F-doped SmFeAsO bulk samples and superconducting wires. As-grown samples were synthesized by the solid-state reaction[11] and ground to a powder in a mortar. The powder was compressed into pellets and sintered at 900 °C for 40 h in a sealed quartz tube. Samples prepared in this way are called resintered samples.

Pellets were also formed by compressing powder and binder together. To obtain the binder, Sm, Fe, and As were mixed in an atomic ratio of 2:3:3 and heated at 850 °C for 10 h in an evacuated silica tube. These stoichiometric sintered materials and stoichiometric $SmF_3$ make up the binder (i.e., the binder contains stoichiometric Sm, Fe, As, and F). 1 mol% as-grown samples and 0.05 mol% binder were powdered and mixed together in a mortar, and then, compressed into pellets. The pellets were sintered at 900 °C for 40 h in a sealed quartz tube. Samples prepared in this way are called resintered samples with binder.

The superconducting wire was fabricated by the *ex-situ* PIT technique. The superconducting core was made from the same raw material as used for the resintered samples with binder. A silver tube inserted into an Fe sheath was used as a sheath material. The inner and outer diameters of the silver tube were 3 and 4 mm, respectively, and the inner and outer diameters of the Fe sheath were 4 and 6.2 mm, respectively. The sheath material packing the raw material was cold rolled into rectangular wire with a cross section of about 2.2 mm$^2$. The wire was sealed in a quartz tube, then annealed at 900 °C for 4 h.

X-ray diffraction (XRD, Rigaku; Rint 2500) with Cu Kα radiation was used to characterize the polycrystalline bulk samples and the superconducting core of the wire, after peeling away the sheath materials. For each bulk sample, the *a*- and *c*-axis lattice parameters were calculated from the positions of Bragg diffraction peaks. Magnetization $M_{mol}$ measurements were also performed for each bulk sample and superconducting core of wire using a Quantum Design magnetic properties measurement system (MPMS). The resistivities $\rho$ of the polycrystalline bulk samples and the superconducting wire with the silver and iron sheath were measured by the standard four-probe technique using Au electrodes. The transport critical currents ($I_c$) of the superconducting wire and the dependence of $I_c$ on magnetic field were measured at the National Institute for Materials Science (NIMS) in Tsukuba, Japan by the standard four-probe technique with a criterion of 1 μV/cm.

Figure 2 shows the XRD patterns of F-doped SmFeAsO powders in the following forms: as-grown, resintered, resintered with binder, and superconducting core of wire. Almost all the diffraction peaks are assigned to the $SmFeAsO_{1-x}F_x$ phase, although there are several minor peaks that are attributed to the SmOF, FeAs, and SmAs phases (impurity phases). The appearance of SmOF phases indicates a partial reduction of fluorine content in the $SmFeAsO_{1-x}F_x$ phase.

The impurity-phase XRD peaks are larger in the resintered sample than in the as-grown sample. The resintered sample with binder exhibits the highest SmOF peak, but the FeAs peak from the resintered sample with binder is slightly smaller than that of the resintered sample. This result indicates that a part of the binder reacts and forms the SmFeAsO$_{1-x}$F$_x$ phase. The lattice parameters along the $a$- and $c$-axes are, respectively 0.3938(4) and 0.8485(6) nm for the as-grown sample, 0.3938(4) and 0.8491(4) nm for the resintered sample, and 0.3930(1) and 0.8474(9) nm for the resintered sample with binder. Because the F content increases with decreasing lattice parameters,[11] the F content of the resintered sample with binder is sufficiently supplemented by adding the binder. The XRD pattern of the superconducting core is very similar to that of the resintered sample with binder, except for a peak attributed to Ag. This result indicates that any reaction between Ag and SmFeAsO$_{1-x}$F$_x$ is negligible.

Figure 3 shows $\rho$–$T$ curves of the samples: as-grown, resintered, resintered with binder, and superconducting wire with silver and iron sheath. The data for the as-grown sample indicate an onset (offset) $T_c$ of 54 K (37 K). The $\rho$–$T$ curve of the resintered sample does not show any clear superconducting transition but shows a finite residual resistivity at 4 K, because the F content in Sm-1111 is reduced below the critical value that is observed in superconducting Sm-1111.[11] However, the $\rho$–$T$ curve of the resintered sample with binder demonstrates an onset (offset) $T_c$ of 55 K (48 K). The observation that the onset and offset $T_c$ of the resintered sample with binder is higher than that of the as-grown sample indicates that resintering with binder suppresses inhomogeneous fluorine substitution in the SmFeAsO$_{1-x}$F$_x$ phase. The $\rho$–$T$ curve for the superconducting wire shows the onset (offset) $T_c$ of 51 K (36 K). Moreover, the superconducting volume fractions (SVF), obtained from $M_{mol}$ measurements, are ~68, ~34, ~104, and ~55 vol. % for the as-grown, resintered, resintered with binder, and superconducting core samples, respectively. The reason why SVF slightly exceeds 100 vol. % in resintered sample with binder can be attributed to the demagnetization effect.[12] The resintered sample with binder has a higher $T_c$ and SVF than does the as-grown sample. Note that resintering with binder not only reduces the drop in $T_c$ but also increases the SVF. The onset $T_c$, offset $T_c$, and SVF are summarized at the bottom of Fig. 1.

Figure 4 shows transport $J_c$ as a function of both increasing and decreasing magnetic flux density $\mu_0H$, and the inset (a) shows $I_c$ at $\mu_0H < 15$ T. The transport $J_c$ attains ~4

kA/cm$^2$ at 4.2 K, which is higher than the other reported values for superconducting SmFeAsO$_{1-x}$F$_x$ wires.[7,8] The maximum transport $J_c$ is observed at $\mu_0 H \sim 0.03$ T for both increasing and decreasing $\mu_0 H$, which is similar to the results obtained for polycrystalline copper-based superconductors.[13] Inset (b) shows applied current versus voltage at 0.03 T. Under both increasing and decreasing $\mu_0 H$, the transport $J_c$ rapidly decreases with increasing field, which is attributed to weak links between grains.

Thus, we demonstrated the first *ex-situ* fabrication of superconducting Sm-1111 wire by using a binder composed of SmF$_3$, Fe arsenide, and Sm arsenide. The introduction of the binder not only reduces the drop in $T_c$ during resintering but also increases the SVF in polycrystalline SmFeAsO$_{1-x}$F$_x$ samples. Although resintering reduces the F content in Sm-1111, adding the binder compensates by sufficiently supplementing the F content. The transport $J_c$ of the superconducting wire with binder attains $\sim$4 kA/cm$^2$ at 4.2 K. The wire demonstrates an onset (offset) $T_c$ of 51 K (36 K). Higher $J_c$ should be achieved by optimizing the binder composition and/or amount in the resintering process.

Acknowledgments

This work was partially supported by the Research Grant of Keio Leading-edge Laboratory of Science & Technology

**Fig.** 1. (Color online) Flowchart for fabricating F-doped SmFeAsO superconducting bulk samples and wire. Superconducting transition temperatures $T_c$, superconducting volume fractions (SVF), and lattice parameters are given for each sample.

**Fig.** 2. (Color online) XRD patterns of F-doped SmFeAsO powders (as-grown, resintered, resintered with binder, and superconducting core of wire). Bottom bars indicate Bragg diffraction positions for $SmFeAsO_{0.92}F_{0.08}$.

**Fig.** 3. (Color online) Resistivity $\rho$ versus temperature $T$ of four F-doped SmFeAsO samples: as-grown, resintered, resintered with binder, and superconducting wire with Ag sheath.

**Fig.** 4. (Color online) Transport critical current density ($J_c$) versus magnetic flux density ($\mu_0 H$) for superconducting F-doped SmFeAsO wire. The maximum transport $J_c$ reaches the value of ~4 kA/cm$^2$ = ~40 MA/m$^2$ at 4.2 K. Inset (a) shows $I_c$ versus $\mu_0 H$. Inset (b) shows applied current versus voltage under $\mu_0 H = 0.03$ T. A slashed lined denotes the criterion. Open (closed) circles denote measurement under increasing (decreasing) $\mu_0 H$.

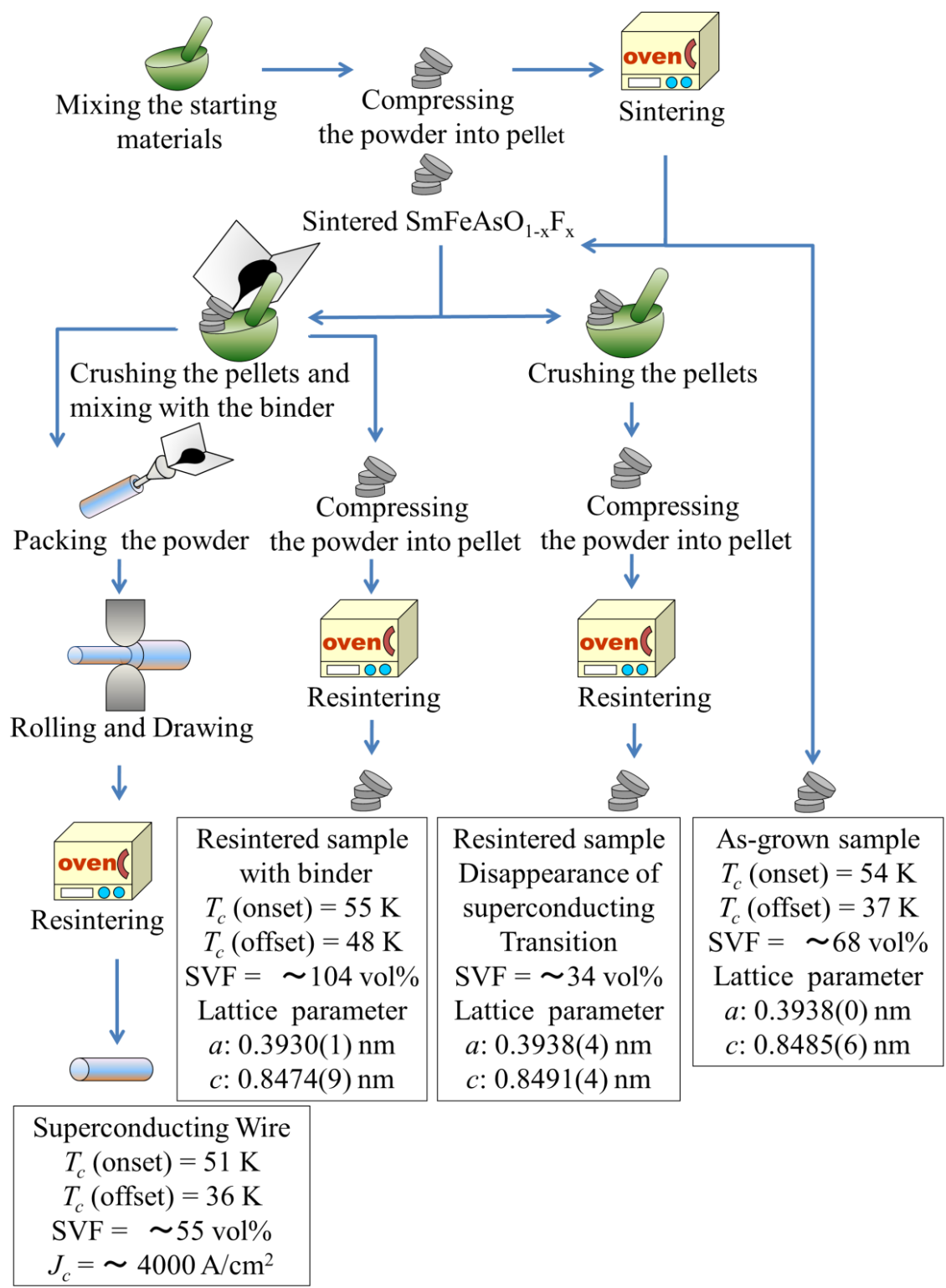

Fig. 1

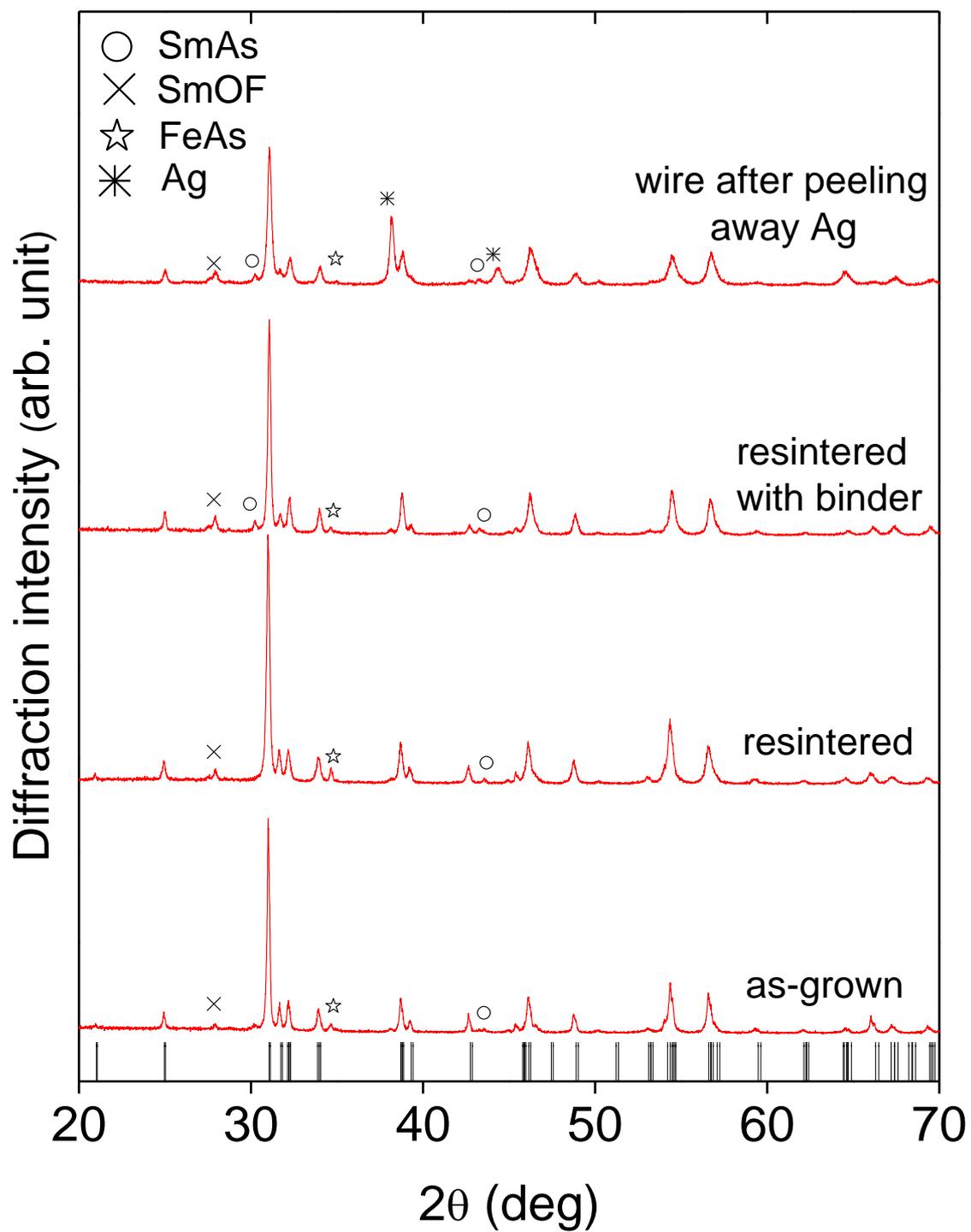

Fig. 2

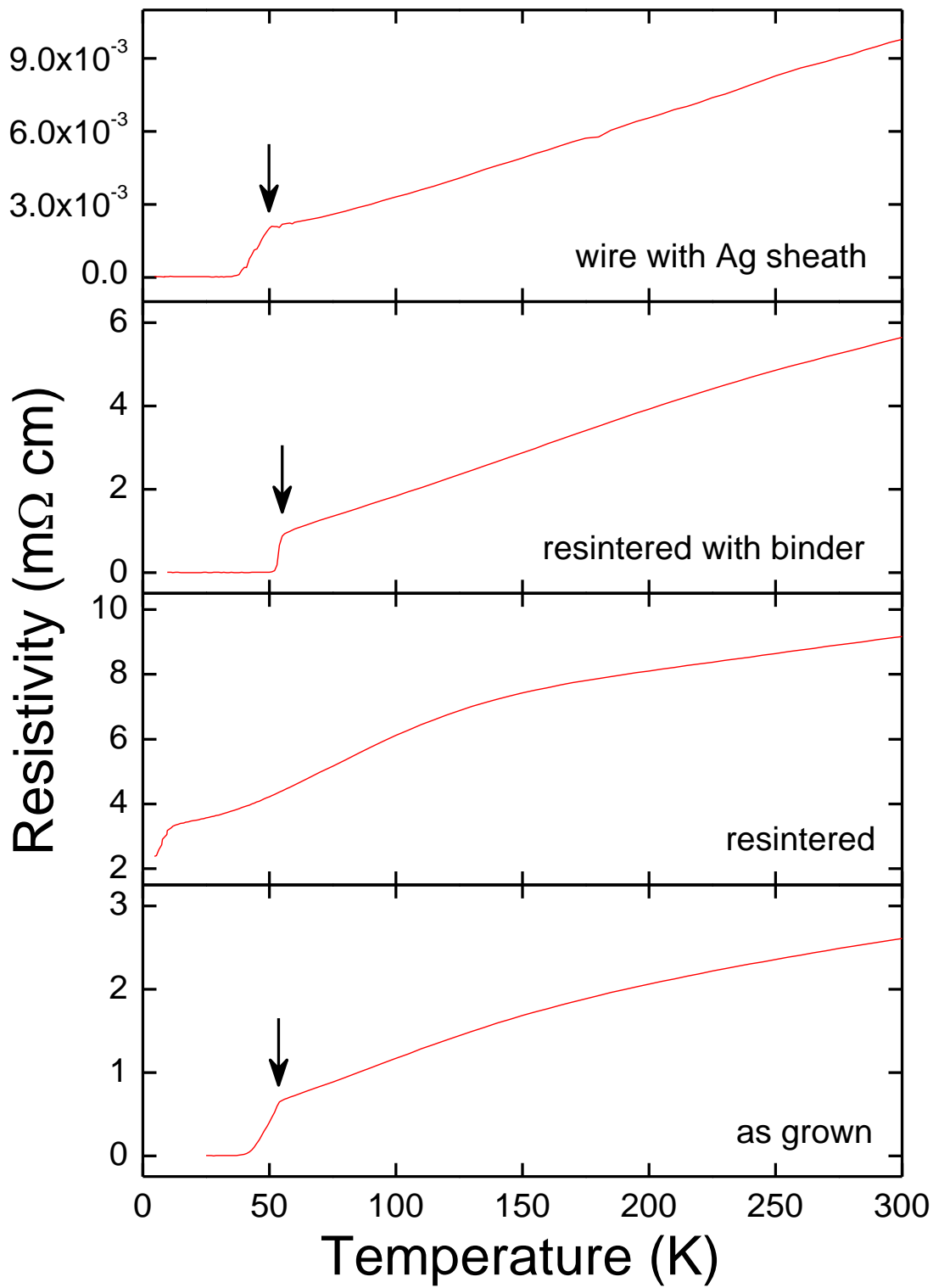

Fig. 3

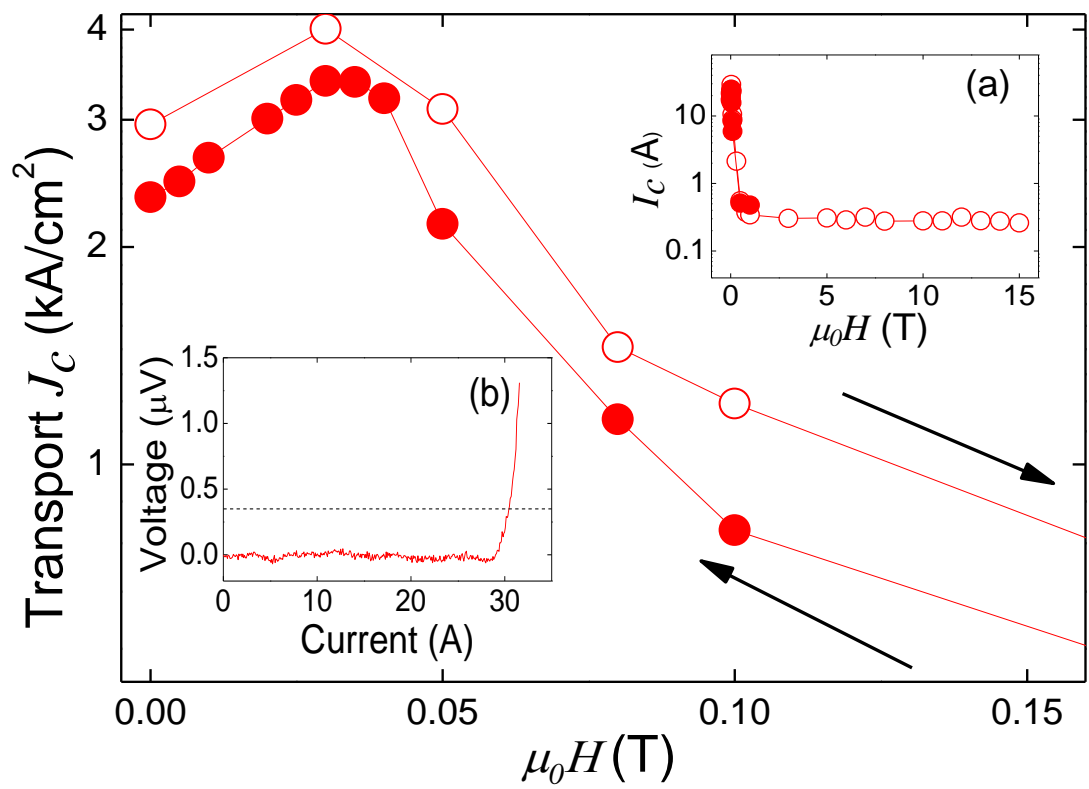

Fig. 4